# A Dynamical Solution of the Triple Asteroid System (45) Eugenia◊

F. Marchis[1,2,3], V. Lainey[3], P. Descamps[3], J. Berthier[3], M. Van Dam[4], I. de Pater[2], B. Macomber[1,2], M. Baek[1,2], D. Le Mignant[5], H. B. Hammel[6], M. Showalter[1], F. Vachier[3]

1. SETI Institute, 515 N. Whisman Road, Mountain View CA 94043 USA
2. University of California at Berkeley, Department of Astronomy, 601 Campbell Hall, Berkeley CA 94720, USA
3. Institut de Mécanique Céleste et de Calcul des Éphémérides, Observatoire de Paris, UMR8028 CNRS, 77 av. Denfert-Rochereau 75014 Paris, France
4. Flat Wavefronts, PO BOX 1060, Christchurch 8140, New Zealand
5. W.M. Keck Observatory, 65-1120 Mamalahoa Hwy, Kamuela, HI 96743, USA
6. Space Science Institute, Boulder, CO 80303, USA

**Abstract:**

We present the first dynamical solution of the triple asteroid system (45) Eugenia and its two moons Petit-Prince (Diameter~7 km) and S/2004 (45) 1 (Diameter~5 km). The two moons orbit at 1165 and 610 km from the primary, describing an almost-circular orbit (e~$6\times10^{-3}$ and e~$7\times10^{-2}$ respectively). The system is quite different from the other known triple systems in the main belt since the inclinations of the moon orbits are sizeable (9° and 18° with respect to the equator of the primary respectively). No resonances, neither secular nor due to Lidov-Kozai mechanism, were detected in our dynamical solution, suggesting that these inclinations are not due to excitation modes between the primary and the moons. A 10-year evolution study shows that the orbits are slightly affected by perturbations from the Sun, and to a lesser extent by mutual interactions between the moons. The estimated $J_2$ of the primary is three times lower than the theoretical one,

---

◊ Based on observations collected at the European Southern Observatory, Chile programs number 270.C-5024, 072.C-0016, 073-C.0062, 077.C.0422 calculated assuming the shape of the primary and an homogeneous interior, possibly suggesting the importance of other gravitational harmonics.

Keywords: Asteroids, Asteroids, dynamic; Satellites, dynamic; Infrared Observations,

## 1. Introduction

The study of asteroids, remnants of the formation of the planets, is key to understanding the past of our solar system. Binary asteroids, asteroids with a satellite, are particularly interesting since they provide a window into the collisional history of our solar system, and are natural laboratories to study surface alteration processes and evolution for asteroids with different sizes, shapes, densities and environments. Since the discovery of the first binary asteroid system, Dactyl orbiting around (243) Ida in 1993 (Chapman et al. 1995), we have learned of ~192 companions of small solar system bodies, including 8 multiple systems composed of more than one companion. The asteroid (87) Sylvia, orbiting in the Cybele part of the main belt, was the first asteroid known to have 2 companions. Its larger moon (87) Sylvia I Romulus was discovered in 2001 (Brown et al. 2001) whereas the closer and smaller moon named (87) Sylvia II Remus was discovered 4 years later (Marchis et al, 2005a). These two moons orbit well inside the Hill sphere of the primary and describe circular, direct, and equatorial orbits (Marchis et al, 2005b). Using adaptive optics technology available on 8-10m class ground-based telescopes, three more triple asteroid systems located in the main belt have been discovered recently: (45) Eugenia (Marchis et al., 2007), (216) Kleopatra (Marchis et al. 2008a), and (93) Minerva (Marchis et al. 2009a). Like (87) Sylvia, we know that these systems have a large primary (D~100-200 km) and two km-sized satellites, but their mutual orbits are not yet defined.

The main-belt asteroid (45) Eugenia is an interesting system in the broad and diverse family of multiple small solar system bodies. A first moonlet officially named

(45) Eugenia I Petit-Prince (hereafter called "Petit-Prince") was discovered using a ground-based adaptive optics system available on the Canada-France-Hawaii Telescope in 1999 (Merline et al. 1999). Its orbit was well constrained using a large set of adaptive optics (AO) data collected from November 1998 to August 2006 based on a Keplerian model (Marchis et al, 2008b). This work revealed that its mutual orbit is direct with respect to the primary and almost circular like those of other binary asteroid systems such as (22) Kalliope, (107) Camilla and (762) Pulcova. However, they reported a significant inclination (~12 deg) for this satellite with respect to the equator of the primary. More recently, Marchis et al. (2007) announced the discovery of a second moon after careful reanalysis of data recorded in February 2004 with the European Southern Observatory (ESO) Very Large Telescope (VLT) Adaptive Optics (AO) system. (45) Eugenia is located roughly in the middle of the asteroid main belt with semi-major axis of about 2.721 AU, eccentricity of 0.083 and 6.61° of orbital inclination.

We propose in this work to estimate the mutual orbits of both known satellites of (45) Eugenia. Additional observations were collected in 2007, shortly after the discovery of S/2004 45 1 (the placeholder name used in the rest of this work will be "Princesse") and are presented in Section 2. In Section 3, we show how we used a dynamical model to derive the osculating elements of the two moon orbits. A long-term temporal analysis of the orbital elements is described and discussed in Section 4.

## 2. Observations

**2.1 Adaptive Optics Data from 1998 to 2007**

Today several adaptive optics systems are available on 8-10m class telescopes. Thanks to real-time correction of atmospheric turbulence effects, images and spectra recorded with these innovative instruments reach an angular resolution close to the diffraction limit of the telescopes. The images also have better contrast, permitting the detection of faint features around bright sources such as satellites of large asteroids. (45) Eugenia is an excellent target for AO observation since i) the asteroid is bright enough at opposition (apparent visual magnitude V~11-12) to be used as a wavefront reference for the AO correction, ii) its proper motion is relatively small (~70 arcsec/hr at its opposition), iii) the primary is not resolved without AO correction providing an excellent source for the wavefront sensor. The data described in this work were essentially collected using two telescopes:

- the Yepun telescope, one of the four ESO 8m-telescopes located at Mount Paranal in Chile, part of the Very Large Telescope. It has been equipped with NaCo which stands for NAOS-CONICA (Lenzen et al. 2003, Rousset et al. 2003), an adaptive optics systems offered since 2003. The near-infrared camera CONICA was used in direct imaging mode with the S13 camera corresponding to a pixel size of 13.27 milli-arcsec (mas) in the Ks band filter (central wavelength 2.18 μm and bandwidth of 0.35 μm).

- the Keck-II 10m telescope located on the top of Mauna Kea, a dormant volcano on the Big Island of Hawaii, equipped with an AO system since 2001 (Wizinowich et al. 2000; van Dam et al. 2004). The data listed in Table 1 were recorded using the near-infrared camera (NIRC2) with a pixel scale of 9.96 mas in the Kp band filter (central wavelength 2.124 μm and bandwidth of 0.336 μm) and the H band filter (central wavelength 1.633 μm and bandwidth of 0.300 μm).

Additional data collected in 1998 using the Canada-France-Hawaii Telescope (CFHT) and its PUEO AO system, retrieved and reanalyzed by Marchis et al. (2008b), are also used for this work.

Data taken from 1998 to 2006 are listed in Table 2b of Marchis et al. (2008b). More recent data taken in 2007 and not used in this previous work are listed in Table 1. These 9 additional observations expand the temporal baseline of the Keplerian model of Petit-Prince and provide 6 astrometric positions of Princesse, allowing us to derive the orbital elements of its orbit.

Reanalysis of the 2004 data collected with the VLT/NACO revealed the presence of a second fainter satellite (~5 km in diameter), closer (~0.4", corresponding to a projected distance of ~600 km) to the primary as shown in Fig. 1 (Marchis et al. 2008b). Additional observations were collected with the Keck AO system in 2007 that confirm the genuineness of this new satellite (Fig. 2). The rate of detection of this small satellite, which is defined as the number of observations with the detected satellite divided by the total number of observations, was quite low (12%), but it increased by 5% since 2007. The recent data taken with the W.M. Keck-II telescope have better sensitivity due to an improvement in quality of the AO system, improving the detection rate of the fainter and closer satellite of (45) Eugenia. Even if the intensity of the inner satellite is close to the intensity of the artifacts in the case of the observations taken on October 19, 2007, it can be easily identified due to its motion around the primary. Additionally, the artifacts visible around the PSF are "ghosts" of the central peak, meaning that they are extended like the resolved primary. In contrast, the moons are circular sources with a full width at half-maximum close to the diffraction limit of the telescope (~45 mas in Kp band).

### 2.2 Data-processing, photometry and astrometry

Each image was basic-processed following the same procedure. The observations of the asteroid were recorded at different locations on the detector (with an individual integration time of ~ 1 min). An estimate of the sky, calculated by a median average of these frames, was subtracted from each individual observation. A flat-field frame and a bad-pixel map of the detector with the relevant filter were calculated using observations of the sky at sunset or sunrise (VLT) or of a uniform light projected on the dome (Keck). Each frame was divided by a normalized flat-field frame to correct for the heterogeneity of the pixel-to-pixel response. The bad pixels in the frames were replaced by the average of their neighborhood pixels. We used the *eclipse* data reduction package to perform the basic data processing (Devillard, 1997). The final frames taken over a time span of less than 10 min were combined into one single average image after applying an accurate shift-and-add process.

We describe in Marchis et al. (2005b, 2008b) how we measure the position of the satellites with respect to the primary. Our algorithm is based on a Moffat-Gauss profile in two dimensions, which is adjusted to fit the position of the satellites and the primary. An estimate of the background due to the residual of the wavefront correction is also added. The astrometric positions relative to the primary in arcsec, labeled X and Y in Table 2, correspond to the projected separation on the celestial sphere between the primary and the satellites. X is positive when the satellite is located to the astronomical East of the primary and Y is positive when it is located to the North. The 1-$\sigma$ errors of the positions, which depend on the brightness of the satellite, its relative position, and the AO correction quality, are estimated to be 20 mas (VLT) and 17 mas (Keck) for Princesse and 9 mas (VLT) and 6 mas (Keck) for Petit-Prince. The 1-$\sigma$ error for the data recorded in 1998 with the CFHT AO system is larger (~70 mas) since the telescope has a modest aperture size of 3.6m. We did not apply any phase correction to correct for the illumination geometry since the observations were taken close to the opposition with an average phase angle of 13.5°. Assuming a spherical shape for the primary, this phase angle would introduce a shift of the centroid of roughly *sin(phase/2)* × $R_{eq}$ *($R_{eq}$* = 108.5 km, the mean radius of the primary), equaling 13 km or 8 mas. This additional error on the astrometric position cannot be corrected since it varies with the shape of the primary which is known to not be spherical, but is not fully characterized (Kaasalainen et al. 2002).

**[insert Table 1]**

**[insert fig 1 &2]**

**[insert Table 2]**

From the Moffat-Gauss profile we also estimate the relative integrated flux between the moon and the primary and derive an estimate of the satellite diameters, assuming that both satellite and primary have the same albedo (Table 2). Two spectroscopic studies in the near-infrared confirm observationally that components of binary asteroid systems have the same color within a 1–$\sigma$ accuracy of 6% for (90) Antiope (Marchis et al. 2009b) and 4% for (22) Kalliope (Laver et al. 2009). This propagates to a relative uncertainty of the estimated size of 3% and 2% respectively, which is considered negligible with respect to the error produced by the imperfection of the AO system estimated after applying our fitting process (1-$\sigma$ error of 20-30%). The diameter of Petit-Prince is estimated to be $7 \pm 2$ km. Princesse, the inner satellite, is slightly smaller with a diameter of $5 \pm 1$ km.

## 3. Mutual Orbits derived from a dynamical model

### 3.1. A better solution with a dynamical model

The geometrical model described in Marchis et al. (2008b) failed to provide a coherent solution that included all the new astrometric positions measured for both satellites of (45) Eugenia. We could only find a solution for Petit-Prince after removing the astrometric positions that were obtained in 1998, 2005 and 2006. Consequently, we used a dynamical model of the system based on the Numerical Orbit and Ephemerides (NOE) code, which has been applied successfully to the Martian system (Lainey et al. 2007), the Uranian system (Lainey 2008), and more recently to the quantification of the tidal accelerations among the Galilean moons of the Jovian system (Lainey et al. 2009). The numerical model is described extensively in the above publications so we will only briefly summarize the concepts. The dynamical system is numerically integrated in a planetocentric frame (centered on the center of mass of (45) Eugenia) with inertial axes corresponding to EME2000 (Earth Mean Equator and Equinox of Epoch J2000) for convenience. **r_i** is the position vector of a moon (i=1, 2 for Petit-Prince and Princesse), so the related equation of motion has the common form of :

$$\ddot{\vec{r}}_i = -\frac{G(m_0 + m_i)\vec{r}_i}{r_i^3} + \sum_{j \neq i} Gm_j \left( \frac{\vec{r}_j - \vec{r}_i}{r_{ij}^3} - \frac{\vec{r}_j}{r_j^3} \right) + G(m_0 + m_i)\nabla_i U_{i0} + \sum_{j \neq i} Gm_j \nabla_j U_{j0} \quad (1)$$

Where $U_{k0}$ introduces the oblateness gravity component of the planet at the position of **r_k**.

$$U_{k0} = -\frac{1}{r_k}\left(\frac{R_{eq}}{r_k}\right)^2 \frac{J_2}{2}\left(3\sin^2\theta_k - 1\right) \quad (2)$$

**[Comment for proof stage:** please use boldface instead of an arrow in Eq. 1 to denote vectors]

$G$ is the gravitational constant, $M$ is the mass of the primary, $r_k$, $\theta_k$, $\lambda_k$ are the spherical coordinates in an equatorial frame ($\theta_k$ is the latitude), $R_{eq}$ = 108.5 km is the mean radius of the primary, defined as a sphere of radius $R_{eq}$ with the same volume as the primary (Marchis et al., 2008b). Since Eq. 1 is integrated in EME2000 frame, one needs to apply two successive rotations of angle $-\pi/2 + \beta$ and $-\lambda - \pi/2$ to transpose the angle from Eugenia's equator frame to EME2000 respectively, where $\lambda$ and $\beta$ are the ecliptic longitude and latitude of the spin axis direction (J2000.0, in radian).

The non-spherical nature of the (45) Eugenia primary will introduce additional non-Keplerian effects, such as the precession of the nodes and the apses of the orbit of the moonlets (Descamps, 2005). The largest of these effects is due to $J_2$ ($=-C_{20}$), the lowest-order gravitational coefficient, which is related to the moments of inertia of the primary by:

$$J_{2theo} = \frac{1}{MR_{eq}^2}\left(C_p - \frac{A_p + B_p}{2}\right) \sim \frac{1}{10R^2}\left(\alpha_p^2 + \beta_p^2 - 2\gamma_p^2\right)$$

(3)

where $A_p$, $B_p$ and $C_p$ are the moments of inertia, and $\alpha_p$, $\beta_p$, $\gamma_p$ are the tri-axial radii of an ellipsoid with approximately the same shape as the primary (Scheeres, 1994). From the Kaasalainen et al. (2002) 3D-shape model, we estimated $J_{2theo}$ ~0.19 assuming a homogeneous distribution of mass in the interior of the primary.

The integrator subroutine included in NOE from Everhart (1985) was chosen for its computational speed and accuracy. A constant step of 0.25 days was used. To increase the numerical accuracy during the fitting procedure we performed forward and backward integrations starting at an initial Julian epoch of 2452980.0 (6 December 2003 at 12:00 UT). This date corresponds to the first high precision measurement of the position of Petit-Prince, recorded using the Keck AO system (see Table 2b in Marchis et al. 2008b).

The model presented in this work takes into account: i) the mass of Eugenia and its $J_2$, ii) the mutual perturbation between Petit-Prince and Princesse, which are assumed to be point masses, iii) the perturbations of the Sun, the 8 major planets of our solar system, Pluto, and Earth's moon, using the planetary ephemeris DE406 (Developed Ephemeris 406). This planetary ephemeris is similar to DE405 but with a longer timeline (Standish et al., 1998).

The following parameters are considered to be already known in the analysis: the mass of Eugenia (2.83E-12 × $M_{sun}$), the mass of Petit-Prince and Princesse assuming a diameter of ~7 km (1.26E-16× $M_{sun}$), and an equivalent diameter for the primary of 217 km (Marchis et al., 2008b). The fitted parameters are the initial conditions of each moon, the primary's $J_2$ and its spin pole's spherical ecliptic coordinates λ and β in EME2000.

During the fitting procedure, time scale and light time corrections for each satellite-observer distance were introduced. The model is determined by the least-squares method (singular value decomposition). As described in Lainey et al. (2007), only the elliptical elements fitting method was considered since a model based on Cartesian coordinates could quickly diverge. Each astrometric position was weighted on the accuracy estimate described in Section 2.2. All unknowns have been fit together at each iteration.

To check the robustness of the solution, we used a random holdout method. We choose to remove randomly 3 observations (about 10% of the whole observation set), and performed a new fit of Petit-Prince solution (including the fit of $J_2$ and the pole coordinates). Such procedure was done 100 times. We observed that in most cases, the new solutions lay inside 3-σ of the nominal solution (and always inside 4-σ). Hence, we conclude that realistic error bars should be 3-σ.

### 3.2 Method and Results

Because the number of astrometric positions for Princesse is limited, the fit is performed in two steps. First, the numerical integration is used to determine the orbital elements of Petit-Prince, the larger moon, for which 38 astrometric positions are known between November 1998 and December 2007 (Table 2). The pole orientation and the $J_2$ are also derived from this step. Six osculating elements (the semi-major axis, eccentricity, inclination, argument of the periapsis, longitude of the ascending node and mean anomaly) of the Petit-Prince orbit (a.k.a the orbital elements of the non-perturbed Keplerian orbit) are listed in Table 3. We list in Table 4 the initial conditions of the dynamical model for this solution.

The orbital elements from Table 3 are in agreement with those determined in Marchis et al (2008b) using a Keplerian model with precession due to the oblateness of the primary. We confirmed the large inclination (~9º) of the Petit-Prince satellite with respect to the Primary's equator, which is atypical for asteroid systems possessing a moon well inside the Hill sphere ($a_{Prince}$~3/100 x $R_{Hill}$) . The gravitational coefficient $J_2$ is 3 times lower than the expected value considering a primary with a 3D-shape derived by Kaasalainen et al. (2002) and a homogenous distribution of material in its interior ($J_{2theo}$~ 0.19). This possibly suggests the importance of the neglected oblateness gravity terms of Eugenia. The pole solution of the primary is well constrained ($\lambda$=122.0° ± 1.2°, $\beta$=-19.2° ± 0.9°) in EME2000 and very close to Kaasalainen et al. (2002) derived pole orientation of (118°, -13°) in EME2000.

The second step consists of fitting the orbit of Princesse using the $J_2$ and pole solution derived previously and based on 6 astrometric positions collected from 14 February 2004 to 19 October 2007 and listed in Table 2. Petit-Prince is included as a perturber with the nominal orbit found in the first step. Table 3 contains the fitted initial osculating elements of both satellites in an EME2000 frame centered on Eugenia at the initial Julian epoch of 2452980.0 (6 December 2003 at 12:00 UT).

**[Insert Table 3]**

This first estimate of the Princesse orbit indicates that the inner satellite is probably more inclined (~18° with respect to the primary's equator) than the outer one while describing still an almost-circular orbit. Princesse orbits very close to the primary at 611 km, corresponding to $5.9 \times R_p$ or $3/200 \times R_{Hill}$. This significant inclination is puzzling since no binary systems with satellites orbiting well inside the Hill sphere and describing circular orbits, such as (22) Kalliope, (107) Camilla, (762) Pulcova (Marchis et al., 2008b), (121) Hermione (Descamps et al. 2009), or triple asteroids such as (87) Sylvia (Marchis et al 2005) and (216) Kleopatra (Descamps et al. 2010) do not display this characteristic. A non-excited orbit should be located nearly in alignment with the equatorial plane of the primary as seen for (87) Sylvia and its two moons (Marchis et al. 2005b). (45) Eugenia is the only multiple main-belt asteroid known so far for which the orbits of the satellites have a significant inclination with respect to the primary's equator. The mutual inclination of the two orbits given by $\cos \phi = \cos i_{Prince} \cos i_{Princesse} + \sin i_{Prince} \sin i_{Princesse} \cos(\Omega_{Prince} - \Omega_{Princesse})$ is $\phi \sim 20 \pm 9°$ (where $i$ is the inclination in EME2000 and $\Omega$ is the longitude of the ascending node). This non-zero mutual inclination suggests that the orbits of (45) Eugenia's satellites are in an excited state like the (136108) Haumea satellite orbits ($\phi_{Haumea} \sim 13.4°$) described in Ragozzine and Brown (2009).

**[insert Table 4]**

Figure 3 shows the residual differences between the computed positions of the numerical model and the measured astrometric positions for each satellite for each epoch of observations. The accuracy of the astrometric positions is estimated in Section 2.2. Due to the small number of observations, no observations were rejected during the least squares inversion process.

**[insert Figure 3]**

## 4. Temporal evolution of the dynamical model

Using the dynamical solution proposed in Section 3.2 we can study the long-term behavior of the orbits of the (45) Eugenia moons. Figures 4 and 5 show the evolution of their osculating elements from 1 January 1995 to 31 December 2004. There are some significant, and periodic, variations of all orbital elements. Table 5 summarizes the characteristics of the osculating elements *a, e ,i,* and *n* over a period of 10 years, corresponding to ~770 and ~2040 revolutions of the satellites Petit-Prince and Princesse, respectively.

It is interesting to notice that the inclination and eccentricity remain stable for both satellites over this long period of time. Their inclination varied by less than 1° and their eccentricity by 0.002 and 0.01 for each moon. This non-zero inclination and non-zero eccentricity suggest that the excited state of the system is not due to satellite-satellite perturbations. A careful analysis using *n*-body code, similar to what was done for (87) Sylvia (Winter et al. 2009) or the (136108) Haumea system (Ragozzine & Brown, 2009) could help identify signs of instabilities. Winter et al. (2009) suggest that the main-belt triple system (87) Sylvia is guaranteed to be stable over 5,000 of years by the oblateness of its primary, even with a $J_2$ as small as 0.02. A similar study for the specific case of the (45) Eugenia triple system, taking into consideration the estimated $J_2$ of 0.06 and its orbit around the sun should be done in the future.

The mean motions are ~1.33 and ~3.5 rad/day suggesting the absence of low-order mean motion resonances in the (45) Eugenia system. No secular resonances (synchronized precession of the periapsis and/or the ascending node) were detected for (45) Eugenia satellites. The two satellites of (87) Sylvia were identified by Winter et al. (2009) to be librating in a secular resonance.

An analysis based on independently disabling the main perturbations in the dynamical model reveals that the variations over a short period (~16 days) and a long period (~840 days) of the inclination of Petit-Prince are mostly due to perturbations from the sun. A small fraction of these variations is introduced by the presence of Princesse as well. The long period variations in the inclination of Princesse are also due to solar perturbations. The short period variations are identical in periodicity with the longitude of the node for the Petit-Prince orbit, suggesting mutual interaction.

We tested if the high inclinations could be due to Lidov-Kozai resonance (Lidov, 1962; Kozai, 1962), a mechanism known to cause periodic exchange between the eccentricity and the inclination. We found that both moons are not locked in Kozai resonance (characterized by the libration of the argument of the periapsis). This is in agreement with Kinoshita & Nakai (2007) since the constant $h$, defined by

$$h = (1 - e^2)\cos^2 i \qquad (4)$$

is larger than the maximum value 0.6 compatible with Kozai resonance (we obtained 0.97 and 0.84 for Petit Prince and Princesse, respectively).

[Insert Table 5]

**[Insert Figure 4 & 5]**

## 5. Discussion and Conclusion

In this work we present the first dynamical model of a triple asteroid system in the main belt. Our dynamical model successfully fits the astrometric positions measured by adaptive optics observations collected over 9 years for (45) Eugenia I Petit-Prince, and 3 years for S/2004 (45) 1 (named in this work "Princesse"). The use of a global dynamical model appeared necessary after we failed adjusting the orbits of the two moons independently as was done for (87) Sylvia in Marchis et al. (2005b). (45) Eugenia is a puzzling multiple system for which the two moons' orbits describe almost circular, but inclined, orbits with respect to the primary's equator. The inclination is significantly larger for the inner satellite, called "Princesse" here ($a$~611 km, $i$~18°), than for Petit-Prince ($a$~1165 km, $i$~9°). No resonances were detected in our dynamical solution, which suggests that these inclinations are not due to excitation modes between the primary and/or the satellites. The long-term evolution of the orbits of the two moons shows that they are affected by the solar perturbations, and to a lesser extent by mutual interaction between the two moons.

The dynamical solution could be improved significantly. The residuals are still relatively important, even if they remain lower than 3-$\sigma$. It is probable that the accuracy of the astrometric position is over-estimated. The primary is well-resolved and irregular in shape, so the centroid position could be shifted by a few mas with respect to the center of mass of the primary. The use of these accuracy estimates as a weight in the inversion process may be also flawed, producing unrealistic error bars on the orbital elements. Finally, and most importantly, the fit of the Princesse orbit could be improved by additional astrometric positions. Several key parameters of the model, for example the mass of Eugenia, are set to a fixed value determined in Marchis et al. (2008b) analysis based on a pure Keplerian model. Our analysis could also be improved by adjusting the larger order coefficients of the gravitational harmonics, $C_{22}$, $C_{21}$, $S_{22}$ and $S_{21}$ for example, thus taking into account the shape and spin of the primary. The fact that the $J_2$ effect is three times lower that the estimated $J_{2theo}$, which was calculated by assuming a homogenous distribution of mass in (45) Eugenia's primary, suggests that the dynamical effect introduced by the irregular shape of the primary is not fully considered in the adjustment due to the lack of astrometric positions. Further astrometric observations of Princesse would help determine if the system is in a resonance which is return, could explain the observed inclinations/mutual inclinations of the satellites.

Despite all these anomalies, we successfully derive a robust dynamical solution of (45) Eugenia and its satellites. A secondary product of this adjustment is the determination of the pole orientation of the primary with an unprecedented accuracy ($\lambda$=122. 0 ± 1.2 deg, $\beta$=-19.2 ± 0.9 deg) in EME2000, which is in agreement with Kaasalainen et al (2002) pole orientation derived from lightcurve inversion and estimated to be (117.9 ± 5.0 deg, -12.7 ± 10.0 deg) in EME2000.

Regular observations of this triple system as has been done for the satellites of major planets such as Mars (Sinclair, 1989) could help constrain the gravitational field of the primary by determining the larger order gravitational harmonics and derive the mass independently. The distribution of material inside large asteroids remains unknown. Because of their low bulk density in comparison to their meteorite analogs (Marchis, 2009c), it has been suggested that most of these multiple asteroids have an interior with large voids, giving a total porosity of 30 to 50% (Britt et al, 2002; Descamps et al. 2007; Descamps et al. 2008; Marchis et al., 2005b; Marchis et al. 2008b,). However, it is unclear if this porosity is due to loosely bound material homogenously distributed within the asteroid or heterogenous voids formed by reaccretion of large irregular fragments during the disruption of the parent asteroid. An accurate and complete dynamical model including a comparison with the theoretical and dynamical coefficients of gravity will help differentiate between these two scenarios.

**Acknowledgements**

FM work was supported in part by the National Science Foundation Science and Technology Center for Adaptive Optics, managed by the University of California at Santa Cruz under cooperative agreement AST 98-76783 and additional support from NASA grant NNX07AP70G. MB & BM were supported by the National Science Foundation under award number AAG-0807468. Parts of these data were obtained with the W.M. Keck Observatory, which is operated by the California Institute of Technology, the University of California, Berkeley and the National Aeronautics and Space Administration. The observatory was made possible by the generous financial support of the W.M. Keck Foundation. The authors extend special thanks to those of Hawaiian ancestry on whose sacred mountain we are privileged to be guests. Without their generous hospitality, none of the observations presented would have been possible.

**Table 1:** Summary of AO observations taken in 2004 and 2007 which reveal the presence of Princesse. This table completes the AO observations listed in Table 2b of Marchis et al (2008).

| ID | Name | Date | UT | total exp (s) | Filter | Telescope |
|----|------|------|------|------|------|------|
| 45 | Eugenia | 14-Feb-04 | 03:41:49 | 300 | Ks | VLT |
| 45 | Eugenia | 15-Feb-04 | 03:30:32 | 300 | Ks | VLT |
| 45 | Eugenia | 16-Feb-04 | 03:42:16 | 300 | Ks | VLT |
| 45 | Eugenia | 19-Oct-07 | 12:05:30 | 900 | Kp | Keck |
| 45 | Eugenia | 19-Oct-07 | 12:54:38 | 360 | H | Keck |
| 45 | Eugenia | 19-Oct-07 | 13:29:59 | 180 | Kp | Keck |

**Table 2:** Astrometric positions of the satellites of (45) Eugenia extracted from the AO images provided by the CFHT in 1998, W.M. Keck-II telescope starting in 2005 (observatory code 568), and the VLT-UT4 telescope in 2004 (observatory code 309). The diameters of the moons are estimated by assuming the same surface composition, hence albedo, between the primary and the moons.

| Date | Time | satellite | codeObs | X arcsec | Y arcsec | Diameter km | elongation deg | phase deg |
|---|---|---|---|---|---|---|---|---|
| 1-Nov-98 | 12:58:39 | Petit-Prince | 568 | 0.461 | 0.609 | - | 135.2 | 13.8 |
| 6-Nov-98 | 14:55:13 | Petit-Prince | 568 | 0.207 | 0.740 | - | 140.5 | 12.4 |
| 6-Nov-98 | 15:23:43 | Petit-Prince | 568 | 0.176 | 0.714 | - | 140.5 | 12.4 |
| 6-Dec-03 | 12:42:32 | Petit-Prince | 568 | -0.638 | 0.435 | 7 | 124.7 | 16.7 |
| 6-Dec-03 | 14:20:10 | Petit-Prince | 568 | -0.673 | 0.365 | 5 | 124.7 | 16.7 |
| 4-Jan-04 | 07:03:54 | Petit-Prince | 309 | -0.827 | -0.012 | 9 | 156.8 | 7.9 |
| 5-Jan-04 | 05:56:03 | Petit-Prince | 309 | -0.118 | -0.709 | 10 | 158.0 | 7.6 |
| 6-Jan-04 | 05:10:22 | Petit-Prince | 309 | 0.788 | -0.410 | 9 | 159.2 | 7.2 |
| 7-Jan-04 | 05:43:15 | Petit-Prince | 309 | 0.522 | 0.522 | 10 | 160.3 | 6.8 |
| 11-Feb-04 | 03:11:20 | Petit-Prince | 309 | -0.822 | -0.192 | 9 | 155.5 | 8.5 |
| 11-Feb-04 | 03:20:05 | Petit-Prince | 309 | -0.826 | -0.197 | 5 | 155.5 | 8.5 |
| 11-Feb-04 | 03:43:53 | Petit-Prince | 309 | -0.834 | -0.210 | 5 | 155.5 | 8.5 |
| 12-Feb-04 | 02:52:35 | Petit-Prince | 309 | 0.081 | -0.725 | 5 | 154.3 | 8.9 |
| 12-Feb-04 | 03:00:54 | Petit-Prince | 309 | 0.084 | -0.727 | 10 | 154.3 | 8.9 |
| 12-Feb-04 | 03:33:25 | Petit-Prince | 309 | 0.115 | -0.723 | 6 | 154.3 | 8.9 |
| 12-Feb-04 | 04:09:55 | Petit-Prince | 309 | 0.131 | -0.720 | 8 | 154.3 | 8.9 |
| 14-Feb-04 | 03:41:49 | Petit-Prince | 309 | 0.317 | 0.658 | 11 | 151.9 | 9.6 |
| 15-Feb-04 | 03:30:32 | Petit-Prince | 309 | -0.714 | 0.426 | 10 | 150.8 | 10.0 |
| 16-Feb-04 | 03:42:16 | Petit-Prince | 309 | -0.622 | -0.456 | 10 | 149.6 | 10.4 |
| 1-Mar-04 | 04:03:50 | Petit-Prince | 309 | -0.672 | -0.342 | 3 | 133.9 | 15.0 |
| 3-Mar-04 | 04:10:27 | Petit-Prince | 309 | 0.826 | 0.052 | 4 | 131.8 | 15.5 |
| 4-Mar-04 | 03:32:56 | Petit-Prince | 309 | 0.139 | 0.626 | 8 | 130.7 | 15.8 |
| 4-Mar-04 | 03:49:08 | Petit-Prince | 309 | 0.140 | 0.627 | 3 | 130.8 | 15.8 |
| 7-Mar-04 | 03:17:19 | Petit-Prince | 309 | 0.517 | -0.508 | 6 | 127.6 | 16.6 |
| 7-Mar-04 | 03:30:42 | Petit-Prince | 309 | 0.533 | -0.498 | 4 | 127.6 | 16.6 |
| 8-Mar-04 | 00:56:15 | Petit-Prince | 309 | 0.770 | 0.172 | 7 | 126.6 | 16.8 |
| 8-Mar-04 | 01:06:29 | Petit-Prince | 309 | 0.763 | 0.185 | 5 | 126.6 | 16.8 |
| 9-Mar-04 | 02:09:22 | Petit-Prince | 309 | -0.105 | 0.624 | 10 | 125.6 | 17.1 |
| 10-Mar-04 | 01:26:36 | Petit-Prince | 309 | -0.771 | 0.089 | 8 | 124.6 | 17.3 |
| 11-Mar-04 | 00:45:46 | Petit-Prince | 309 | -0.289 | -0.560 | 3 | 123.6 | 17.5 |
| 17-Jul-05 | 07:25:44 | Petit-Prince | 568 | 0.328 | 0.095 | 8 | 116.2 | 21.2 |
| 3-Aug-06 | 14:07:40 | Petit-Prince | 568 | 0.149 | -0.636 | 4 | 100.3 | 23.4 |
| 9-Sep-07 | 14:31:52 | Petit-Prince | 568 | 0.412 | 0.229 | 8 | 85.4 | 20.0 |
| 9-Sep-07 | 15:20:50 | Petit-Prince | 568 | 0.420 | 0.265 | 7 | 85.4 | 20.0 |
| 19-Oct-07 | 12:05:30 | Petit-Prince | 568 | -0.564 | -0.177 | 9 | 119.7 | 17.2 |
| 19-Oct-07 | 12:54:38 | Petit-Prince | 568 | -0.558 | -0.199 | 9 | 119.7 | 17.2 |
| 19-Oct-07 | 13:29:59 | Petit-Prince | 568 | -0.567 | -0.226 | 9 | 119.7 | 17.2 |
| 13-Dec-07 | 07:59:08 | Petit-Prince | 568 | 0.433 | 0.731 | 6 | 170.1 | 3.3 |
| 14-Feb-04 | 03:41:49 | Princesse | 309 | 0.127 | -0.326 | 6 | 155.5 | 8.5 |
| 15-Feb-04 | 03:30:32 | Princesse | 309 | -0.216 | 0.261 | 6 | 150.8 | 10.0 |
| 16-Feb-04 | 03:42:16 | Princesse | 309 | 0.385 | -0.231 | 5 | 149.6 | 10.4 |
| 19-Oct-07 | 12:05:30 | Princesse | 568 | -0.089 | -0.286 | 6 | 119.7 | 17.2 |
| 19-Oct-07 | 12:54:38 | Princesse | 568 | -0.073 | -0.276 | 6 | 119.7 | 17.2 |
| 19-Oct-07 | 13:29:59 | Princesse | 568 | -0.065 | -0.295 | 3 | 119.7 | 17.2 |

**Table 3:** Osculating elements of Petit-Prince and Princesse orbits EME2000 at the initial epoch $t_0$ = Julian date 2452980.0 (6 December 2003 at 12:00 UT). In the case of Petit-Prince, the $J_2$ and the orientation of the pole of the primary were also adjusted. All error bars are the formal 3-$\sigma$ error bars which were derived by a random holdout method (see text in Section 3.1).

| Component | Parameter | Value | Error | Units |
|---|---|---|---|---|
| Eugenia | Mass | 2.83 | set | $10^{-12} \times M_{sun}$ |
| | Equivalent Diameter | 217 | set | km |
| | $J_2$ | 0.060 | 0.002 | |
| | Pole solution in ecliptic EME2000 | λ=122.0 | 1.2 | degrees |
| | | β=-19.2 | 0.9 | |
| Petit-Prince | Mass | 1.26 | set | $10^{-16} \times M_{sun}$ |
| | Semimajor axis | 1164.51 | 0.03 | km |
| | Eccentricity | 0.006 | 0.024 | |
| | Inclination in EME2000 | 107.6 | 2.1 | degrees |
| | Longitude of the Ascending node | 202.5 | 0.6 | degrees |
| | Argument of periapse | 138 | 48 | degrees |
| | Mean anomaly | 5 | 48 | degrees |
| Princesse | Mass | 1.26 | set | $10^{-16} \times M_{sun}$ |
| | Semimajor axis | 610.8 | 0.3 | km |
| | Eccentricity | 0.069 | 0.015 | |
| | Inclination in EME2000 | 127.0 | 30 | degrees |
| | Longitude of the Ascending node | 210 | 3 | degrees |
| | Argument of periapse | 95 | 63 | degrees |

| Mean anomaly | -186 | 54 | degrees |

**Table 4:** Initial position and velocity vectors in EME2000 frame centered on Eugenia at the initial epoch in Julian date 2452980.0 (6 December 2003 at 12:00 UT) corresponding to the best fit of parameters shown in Table 3.

| **Princesse** | | | |
|---|---|---|---|
| x,y,z (in km) | 198.698 | - 341.315 | -519.467 |
| $v_x, v_y, v_z$ (in km/day) | -1732.40 | -1000.53 | -23.1016 |
| **Petit-Prince** | | | |
| x,y,z (in km) | 772.798 | 548.518 | 664.893 |
| $v_x, v_y, v_z$ (in km/day) | 1013.81 | 11.8257 | -1186.65 |

**Table 5:** Evolution of the osculating elements *a, e, i, n* of Petit-Prince and Princesse reckoned to Eugenia's equatorial frame from January 1 1995 to December 31 2004.

|  | **Petit-Prince** | **Princesse** |
|---|---|---|
| **Semi-major axis** | | |
| <*a*> in km | 1164.6 | 611.1 |
| min,max (*a*) | 1164.4, 1164.8 | 610.8, 611.6 |
| 1-σ (*a*) | 0.1 | 0.2 |
| **Mean motion** | | |
| <*n*> in rad/day | 1.3322 | 3.5047 |
| min, max (*n*) | 1.3318, 1.3326 | 3.5008, 3.5077 |
| 1-σ (*n*) | 0.0002 | 0.0016 |
| **Eccentricity** | | |
| <*e*> | 0.0051 | 0.0708 |
| min, max (*e*) | 0.0040, 0.0062 | 0.0682, 0.0738 |
| 1-σ (*e*) | 0.0006 | 0.0018 |
| **Inclination** | | |
| <*i*> in deg | 9.22 | 18.10 |
| min, max (*i*) | 8.97, 9.35 | 17.98, 18.19 |
| 1-σ (*i*) | 0.10 | 0.05 |
| **Mean rate for the node** | | |
| <dΩ/dt> in deg/day | -0.006 | -0.550 |
| min, max (dΩ/dt) | -0.068, -0.050 | -0.563, -0.537 |
| 1-σ (dΩ/dt) | 0.004 | 0.006 |
| **Mean rate for periapsis** | | |
| <dω/dt> in deg/day | -0.003 | 1.0 |
|  | -2.16, -2.65 | 0.4, 1.6 |
| min, max (dω/dt) | 1.42 | 0.4 |
| 1-σ (dω/dt) | | |

**Figure 1:** 2004 observations recorded with the VLT-UT4 and its AO system NACO. The two moons of (45) Eugenia, Petit-Prince and "Princesse", labeled by horizontal and vertical arrows respectively, can be seen in the low intensity level of the image. The center of the image shows the shape if the primary after sharpening by the deconvolution process.

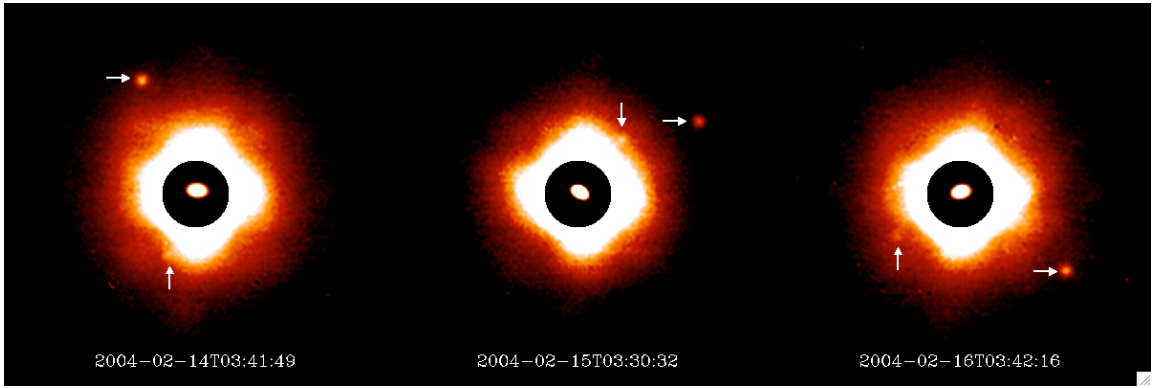

**Figure 2:** 2007 observations recorded with the W.M. Keck-II telescope and its AO system. The two moons of (45) Eugenia, Petit-Prince and "Princesse" are labeled with horizontal and vertical arrows respectively. The center of the image shows the shape if the primary after a sharpening by the deconvolution process using AIDA.

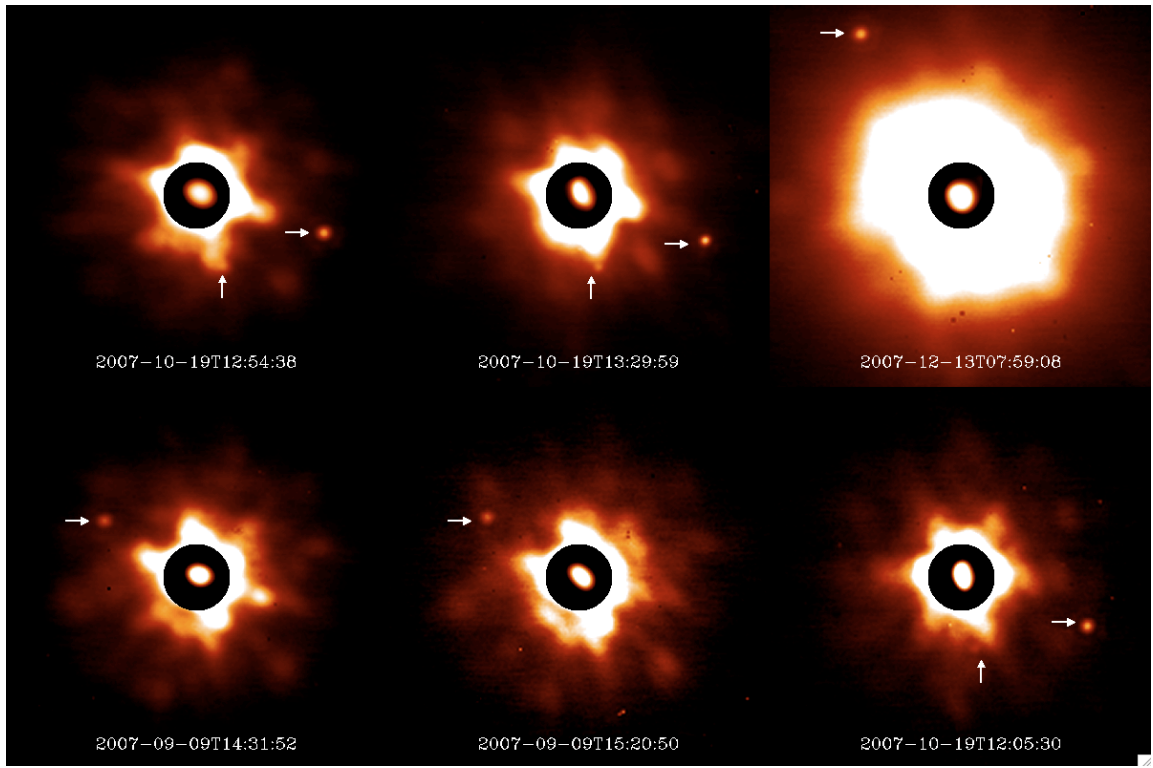

**Figure 3:** Difference, in arcsec, of the measured astrometric positions and the positions generated by the numerical model of each moon of (45) Eugenia. The error is less than 3-σ of the accuracy of the astrometric estimate, confirming the robustness of the orbital analysis.

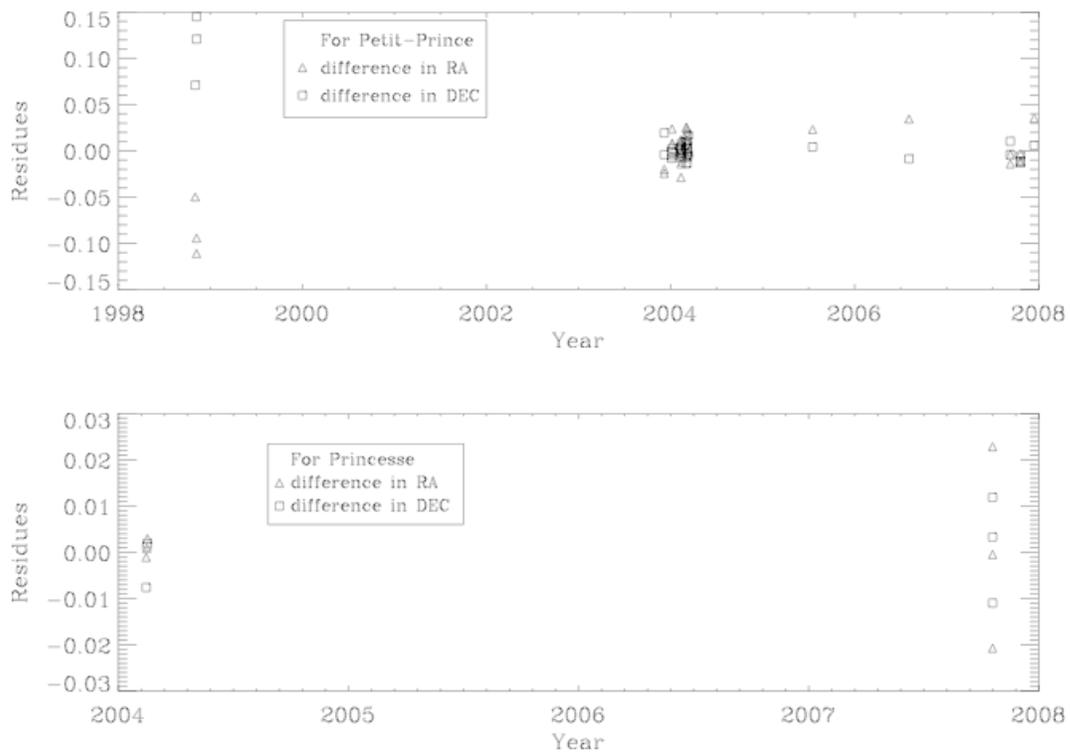

**Figure 4:** Variation of the osculating elements: *the* semi-major axis in km *a,* the eccentricity *e,* and the inclination *i,* of the mutual orbits in Eugenia's equator frame for for Petit-Prince (bottom bold black line) and Princesse (top thin gray line) over 10 yrs starting on 1 January 1995.

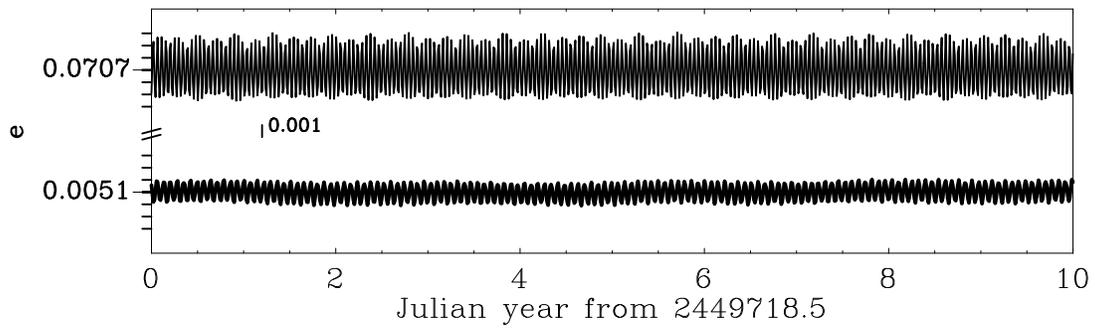

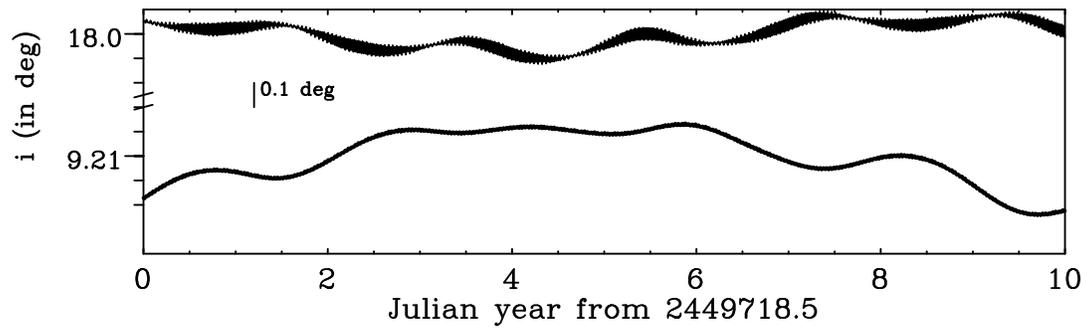

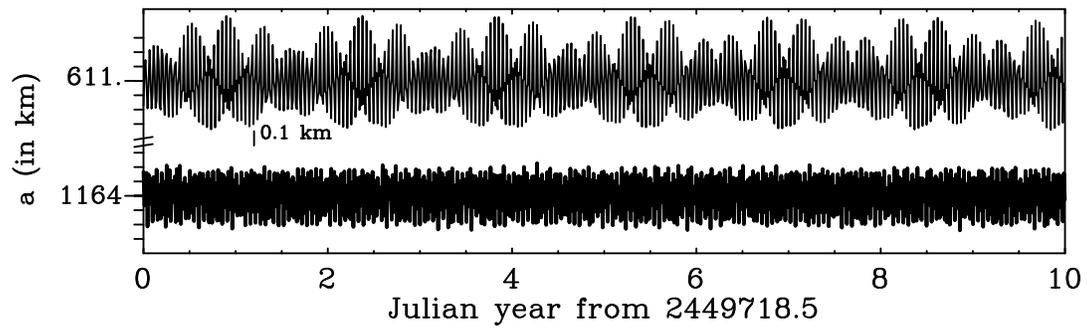

**Figure 5:** Variation of the osculating elements, mean motion ($n = 2\pi/Period$), longitude of the ascending node ($\Omega$), argument of the periapsis ($\omega$) for Petit-Prince (bold black line) and Princesse (thin gray line) over 10 yrs of observations starting on 1 January 1995.

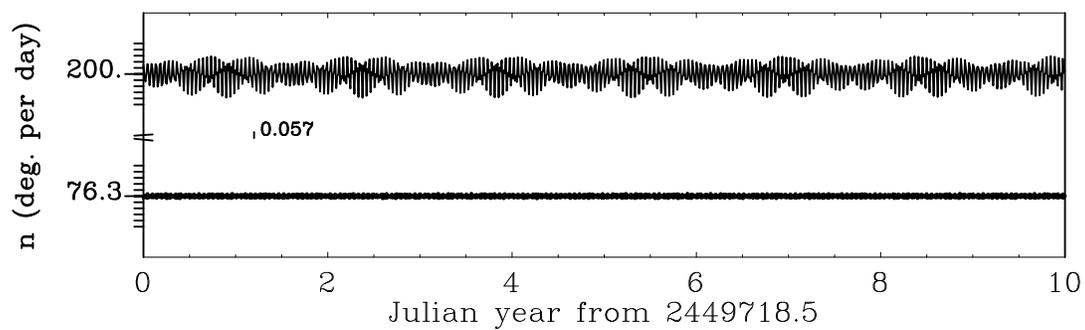
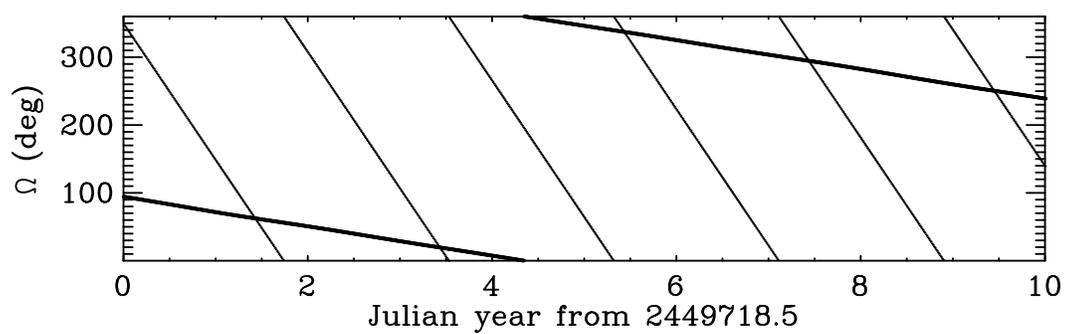
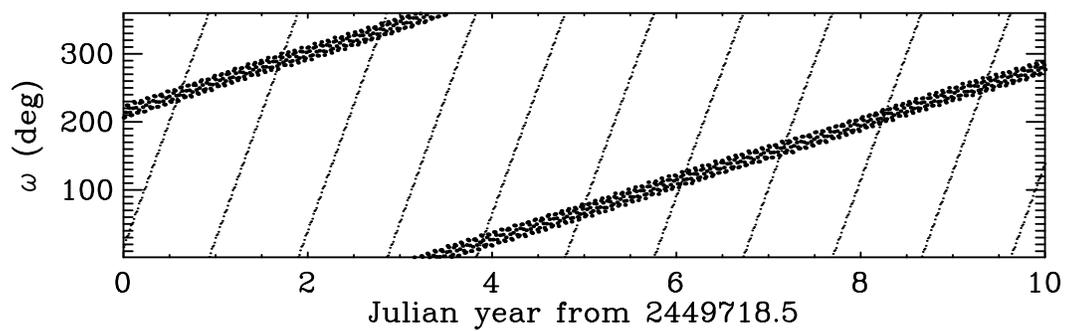